\documentclass[aps,prl,reprint,groupedaddress]{revtex4-1} 
\usepackage[colorlinks=true,urlcolor=blue]{hyperref}
\usepackage{amssymb,amsmath,amsfonts}
\usepackage{epsfig}
\usepackage{graphicx}
\usepackage{epstopdf}

\def\clock{{\count0=\time
           \divide\count0 60
           \ifnum\count0<10 0\fi\the\count0
           \multiply\count0 -60 \advance\count0 \time
           :\ifnum\count0<10 0\fi \the\count0
         }}
\newcommand{\timestamp}{{\small\vbox{\hbox{\tt\jobname.tex}
\hbox{\the\day/\the\month/\the\year, \clock}}}}

\newcommand{\be}{\begin{eqnarray}}
\newcommand{\ee}{\end{eqnarray}}
\newcommand{\beq}{\begin{eqnarray}}
\newcommand{\eeq}{\end{eqnarray}}

\newcommand{\beqa}{\begin{eqnarray}}
\newcommand{\eeqa}{\end{eqnarray}}

\let\oldsqrt\sqrt
\def\sqrt{\mathpalette\DHLhksqrt}
\def\DHLhksqrt#1#2{%
\setbox0=\hbox{$#1\oldsqrt{#2\,}$}\dimen0=\ht0
\advance\dimen0-0.2\ht0
\setbox2=\hbox{\vrule height\ht0 depth -\dimen0}%
{\box0\lower0.4pt\box2}}

\begin{document}

\title{Torsional Newton--Cartan Geometry and Lifshitz Holography}

\author{Morten H. Christensen$^1$, Jelle Hartong$^1$, Niels A. Obers$^1$ and B. Rollier$^2$}
\email[]{mchriste@fys.ku.dk, hartong@nbi.dk, obers@nbi.dk, B.R.Rollier@uva.nl}
\affiliation{$^1$The Niels Bohr Institute, University of Copenhagen}
\address{Blegdamsvej 17, 2100 Copenhagen \O}
\affiliation{$^2$Institute for Theoretical Physics, University of Amsterdam}
\address{Science Park 904, Postbus 94485, 1090 GL Amsterdam, The Netherlands}



\begin{abstract}

We obtain the Lifshitz UV completion in a specific model for $z=2$ Lifshitz geometries. We use a vielbein formalism which enables identification of all the sources as leading components of well-chosen bulk fields. We show that the geometry induced from the bulk onto the boundary is a novel extension of Newton--Cartan geometry with a specific torsion tensor. We explicitly compute all the vevs including the boundary stress-energy tensor and their Ward identities. After using local symmetries/Ward identities the system exhibits 6+6 sources and vevs. The FG expansion exhibits, however, an additional free function which is related to an irrelevant operator whose source has been turned off. We show that this is related to a second UV completion.




 


\end{abstract}


\maketitle


\noindent\textbf{Introduction.} The field of gauge/gravity duality has witnessed the rise of a new paradigm in the form of the AdS/CMT correspondence. By and large the focus of this line of research has been on systems that are described by some strongly coupled CFT. There are however many systems that in the neighborhood of some quantum critical regime are well-described by a scale invariant theory with dynamical exponent $z>1$. Such systems are typically invariant under the Lifshitz symmetry group that contains $z$-dependent scale transformations, space-time translations and spatial rotations. The space-time that has these symmetries as its isometries is known as the Lifshitz space-time and the goal is to develop holographic techniques for space-times that become Lifshitz in some asymptotic region \cite{Kachru:2008yh,Taylor:2008tg}. 

An important objective in this endeavour is the computation of 
 quantities like boundary correlators in various asymptotically Lifshitz backgrounds such as Lifshitz black holes.
This requires that we have an understanding of the boundary geometry and that we control the solutions to the equations of motion near the boundary (in the spirit of Fefferman--Graham (FG) expansions) in order to apply the usual holographic correspondence 
in which the boundary values of the bulk fields act as sources in the dual field theory partition function. 
So far, one has almost exclusively studied such questions in the context of the massive vector model 
\cite{Ross:2009ar,Ross:2011gu,Baggio:2011cp,Mann:2011hg,Griffin:2011xs} because it is simple in matter content (metric and massive vector) and rich in solutions (all values of $z>1$ can be accounted for). However the analogue FG expansions have proven hard to obtain and consequently many features of Lifshitz holography have remained elusive.

The purpose of this letter is to solve this problem in a specific $z=2$ Lifshitz model, allowing to explicitly address the holographic dictionary, including the corresponding boundary geometry, the identification of sources+vevs and computation of
Ward identities. While being the first example that allows such detailed analysis, it is expected that some of the results and methods, in particular the appearance of TNC (torsional Newton-Cartan) geometry on the boundary,  
may serve as important input in the treatment of more general Lifshitz models. 
The advantage of focussing on $z=2$ is that one avoids having to derive the full space of asymptotic solutions to the 
equations of motion since a $z=2$ Lifshitz space-time can be uplifted to an asymptotically AdS space-time in one dimension higher \cite{Balasubramanian:2010uk,Costa:2010cn}.
This observation has led one to look for Lagrangians that in 4 dimensions admit $z=2$ Lifshitz solutions and that can be uplifted to 5 dimensions where they admit asymptotically AdS solutions \cite{Donos:2010tu,Cassani:2011sv,Chemissany:2011mb}. The idea is then to construct the FG expansions of the solutions in 5 dimensions and to reduce this to 4 dimensions. A first step in this direction was taken in \cite{Chemissany:2012du} where the focus was on deriving the counterterms in 4 dimensions.

\noindent\textbf{The Model.} More concretely we study the following model
\begin{equation}\label{eq:actionmodel}
S_{\text{ren}} =\frac{1}{2\kappa_5^2}\int_{\mathcal{M}}d^5x\hat{\mathcal{L}}_{(5)}+\frac{1}{\kappa_5^2}\int_{\partial\mathcal{M}}d^4x\sqrt{-\hat h}\hat K+S_{\text{ct}}\,,
\end{equation}
where the 5D bulk Lagrangian $\hat{\mathcal{L}}_{(5)}$ reads
\begin{equation}
\hat{\mathcal{L}}_{(5)}=\sqrt{-\hat g}\left(\hat R+12-\frac{1}{2}\partial_{\hat\mu}\hat\phi\partial^{\hat\mu}\hat\phi-\frac{1}{2}e^{2\hat\phi}\partial_{\hat\mu}\hat\chi\partial^{\hat\mu}\hat\chi\right)\,,
\end{equation}
and where the counterterm action (we use minimal subtraction) is 
\begin{eqnarray}
S_{\text{ct}} & = &
\frac{1}{\kappa_5^2}\int_{\partial\mathcal{M}}d^4x\left(-\frac{1}{4}\hat{\mathcal{L}}_{(4)}
+\sqrt{-\hat h}\hat{\mathcal{A}}\log r\right)\,,\label{eq:Sct1}
\end{eqnarray}
in which $\hat{\mathcal{L}}_{(4)}$ is the 4D version of $\hat{\mathcal{L}}_{(5)}$ with metric $\hat h_{\hat a\hat b}$ defined via
$d\hat{s}^2= r^{-2}dr^2+\hat h_{\hat a\hat b}dx^{\hat a}dx^{\hat b}$ and where $\hat{\mathcal{A}}$ is an anomaly counterterm whose explicit form we do not need. Our notation is as follows: 5D quantities/indices are hatted while 4D
quanties do not have a hat. Further, $a,b$-type indices refer to the boundary spacetime and
underlined indices denote tangent space. 

The FG expansion of the solution to the equation of motion near the boundary at $r=0$ is given by \cite{Papadimitriou:2011qb,Chemissany:2012du}
\begin{equation}
\hat X =  r^{-\hat\lambda} \left( \hat X_{(0)} + r^2 \hat X_{(2)}+r^4\log r \hat X_{(4,1)} + r^4\hat X_{(4)}+\ldots 
\right)  \, , \label{eq: sol phi}
\end{equation}
where $\hat X$ stands for any of the fields $\hat h_{\hat a\hat b}$, $\hat \phi$, $\hat \chi$ and where $\hat\lambda=2,0,0$, respectively. The $\hat X_{(4)}$ coefficients contain the vevs, i.e. symbolically
\begin{equation}
\hat h_{(4)\hat a\hat b}  \supset \frac{\hat t_{\hat a\hat b}}{2} , \;\;
\hat \phi_{(4)} \supset -  \frac{ \langle \hat{\mathcal{O}}_{\hat\phi} \rangle }{2} , \;\;
e^{2 \hat \phi_{(0)}} \hat \chi_{(4)} \supset -    \frac{ \langle \hat{\mathcal{O}}_{\hat\chi} \rangle}{2}  \,.
\end{equation}
Here, the vevs are defined in the usual way as
\begin{equation}
\left( \hat t_{\hat a\hat b},\langle \hat{\mathcal{O}}_{\hat\phi} \rangle , \langle \hat{\mathcal{O}}_{\hat\chi} \rangle \right)
=  \kappa \left(  \frac{2\delta}{\delta\hat h_{(0)}^{\hat a\hat b}} ,
\frac{\delta }{\delta\hat \phi_{(0)}}, \frac{\delta}{\delta\hat \chi_{(0)}} \right)
 S_{\text{ren}}^{\text{os}} 
 \end{equation}
where $\kappa \equiv- \kappa_5^2 (-\hat h_{(0)})^{-1/2}$ and where $S_{\text{ren}}^{\text{os}}$ is the on-shell version of \eqref{eq:actionmodel} and they satisfy the following Ward identities
\begin{eqnarray}
\hat t^{\hat a}_{\;\;\hat a} &=& \hat{\mathcal{A}}_{(0)}\,,\label{eq:tracet}\\
\hat\nabla_{(0)\hat a}\hat t^{\hat a}{}_{\hat b} &=& -\langle \hat{\mathcal{O}}_{\hat\phi} \rangle\partial_{\hat b}\hat\phi_{(0)} -\langle \hat{\mathcal{O}}_{\hat\chi} \rangle\partial_{\hat b}\hat\chi_{(0)} \,,\label{eq:divt}
\end{eqnarray}
with $\hat{\mathcal{A}}_{(0)}$ the leading term in $\hat{\mathcal{A}}$. Here we will not need the explicit form of $\hat{\mathcal{A}}_{(0)}$ (for this see \cite{Christensen:2013rfa}).

\noindent\textbf{Reduction and Sources.} 
As shown in \cite{Chemissany:2012du} there is a subet of the full set of solutions of the 5D theory that can be reduced to 4D asymptotically Lifshitz geometries. This subset involves a Scherk--Schwarz reduction in which the 5D axion has the form $\hat\chi=ku+\chi$ with $\chi$ a 4D axion and where $u\sim u+2\pi L$ parametrizes the reduction circle. All other fields satisfy the ordinary Kaluza--Klein (KK) ansatz for a circle reduction. We write the 4D Einstein frame metric in vielbein basis as follows
\begin{equation}
ds^2=e^\Phi \frac{dr^2}{r^2}+\left(-e^{\underline{t}}_ae^{\underline{t}}_b+\delta_{\underline{i}\underline{j}}e^{\underline{i}}_ae^{\underline{j}}_b\right)dx^adx^b\,,
\end{equation}
where $\underline{i}=1,2$. Defining 5D vielbeins as follows $\hat h_{\hat a\hat b }= -\hat e^{+}_{\hat a}\hat e^{-}_{\hat b}-\hat e^{+}_{\hat b}\hat e^{-}_{\hat a}+\delta_{\underline{i}\underline{j}}\hat e^{\underline{i}}_{\hat a}\hat e^{\underline{j}}_{\hat b}$, the KK ansatz reads 
\begin{eqnarray}
\hat e^{+}_u & = & -\hat e^{-}_u=\frac{1}{\sqrt{2}}e^\Phi\,,\;\;\; 
\hat e^{\underline{i}}_u  =  0\,, \;\;\; 
\hat e^{\underline{i}}_a = e^{-\Phi/2}e^{\underline{i}}_a \,, 
\label{eq:framechoice1} \\
\hat e^{\pm}_a & = &\pm  \frac{1}{\sqrt{2}}e^\Phi\left(A_a \pm e^{-3\Phi/2}e^{\underline{t}}_a\right)\,,
\end{eqnarray}
where $A_\mu$ is the 4D bulk gauge field with $A_r=0$. For $k\neq 0$ the set of 4D solutions consists of two branches: i). those that are asymptotically Lifshitz and ii). those that are asymptotically conformally AdS \cite{Kanitscheider:2008kd} with hyperscaling exponent $\theta=-1$ \cite{Charmousis:2010zz,Huijse:2011ef}. From a 5D perspective the former has the constraint $\hat h_{(0)uu}=0$ and the latter has $\hat h_{(0)uu}>0$. When we take $\hat h_{(0)uu}=0$ the leading behavior of the bulk fields reads
\begin{eqnarray}
A_a+e^{-3\Phi/2}e^{\underline{t}}_a & = & 2r^{-2}e^{-\Phi_{(0)}/2}\tau_{(0)a}+\ldots\,,  \label{etexp} \\
A_a-e^{-3\Phi/2}e_{a}^{\underline{t}} & = & A_{(0)a}+\ldots\,,\\
e^{\underline{i}} & = & r^{-1}e^{\Phi_{(0)}/2}e_{(0)a}^{\underline{i}}dx^a+\ldots\,, \label{eiexp}
\end{eqnarray}
and the constraint $\hat h_{(0)uu}=0$ becomes
\begin{equation}\label{eq:constraint4D2}
e^{2\Phi_{(0)}}=\hat h_{(2)uu}=-\frac{1}{4}\left(\varepsilon_{(0)}^{abc}\tau_{(0)a}\partial_b \tau_{(0)c}\right)^2+\frac{k^2}{4}e^{2\phi_{(0)}}\,,
\end{equation}
where $\varepsilon_{(0)}^{abc}=e_{(0)}^{-1}\epsilon^{abc}$ with $\epsilon^{abc}$ the Levi-Civita symbol and $e_{(0)}=\text{det}(\tau_{(0)a},e_{(0)a}^{\underline{i}})$ the boundary determinant. We have thus identified the following sources: $\tau_{(0)a},e_{(0)a}^{\underline{i}},A_{(0)a},\Phi_{(0)},\phi_{(0)},\chi_{(0)}$.

\noindent\textbf{Boundary Geometry.} For the leading components in the expansion of the inverse vielbeins we introduce $v_{(0)}^a$ and $e_{(0)\underline{i}}^{a}$ defined via
\begin{eqnarray}
\tau_{(0)a}v_{(0)}^a & = & -1\,,\; \; \; \tau_{(0)a}e_{(0)\underline{i}}^a  =  0\,,\\
e_{(0)a}^{\underline{i}}v_{(0)}^a & = & 0\,,\; \; \; e_{(0)a}^{\underline{i}}e_{(0)\underline{j}}^a = \delta^{\underline{i}}_{\underline{j}}\,.
\end{eqnarray}
We then transform from frame to coordinate components on the boundary using 
$X_{(0)\underline{t}}=-X_{(0)a}v_{(0)}^a$, $ X_{(0)\underline{i}}=X_{(0)a}e_{(0)\underline{i}}^a$
and  $ X_{(0)a}=X_{(0)\underline{t}}\tau_{(0)a}+X_{(0)\underline{i}}e_{(0)a}^{\underline{i}}$. In order to understand better the metric structure of the boundary we perform a bulk local Lorentz transformation that leaves $dr/r$ invariant and expand this for small $r$ so that we can read off the induced transformation on the boundary vielbeins. The result is that they transform under the contracted Lorentz group consisting of rotations and Galilean boosts as
\begin{eqnarray}
\hspace{-.5cm}\delta \tau_{(0)a} & = & 0\,,\quad\delta e^{\underline{i}}_{(0)a} = \tau_{(0)a}\Lambda_{(0)}^be_{(0)b}^{\underline{i}}+\Lambda_{(0)\underline{j}}^{\underline{i}}e^{\underline{j}}_{(0)a}\,,\label{eq:booste0_a}\\
\hspace{-.5cm}\delta v^a_{(0)} & = & \Lambda_{(0)}^a\,,\quad\delta e^a_{(0)\underline{i}} = -\Lambda_{(0)\underline{i}}^{\underline{j}}e_{(0)\underline{j}}^a\,,\label{eq:rotatione0^a}
\end{eqnarray}
where $\Lambda_{(0)}^a=\Lambda_{(0)}^{\underline{i}}e_{(0)\underline{i}}^a$ corresponds to local boosts and
$\Lambda_{(0)\underline{i}\underline{j}}=-\Lambda_{(0)\underline{j}\underline{i}}$ to local $SO(2)$ rotations. The flat index $\underline{i}$ can be raised and lowered with $\delta_{\underline{i}\underline{j}}$.
There are two degenerate metrics invariant under the local tangent space group: $\tau_{(0)a}\tau_{(0)b}$ and $\Pi_{(0)}^{ab}=\delta^{\underline{i}\underline{j}}e^{a}_{(0)\underline{i}}e^{b}_{(0)\underline{j}}$. Further the boundary determinant $e_{(0)}$ is an invariant as well. Hence we cannot raise and lower indices.

Let us construct covariant derivatives $\mathcal{D}^T_{(0)a}$ that transform covariantly under \eqref{eq:booste0_a}
\begin{eqnarray}
\hspace{-.5cm}\mathcal{D}^T_{(0)a} \tau_{(0)b} & = & \nabla^T_{(0)a}\tau_{(0)b}=0\,,\label{eq:nablatau}\\
\hspace{-.5cm}\mathcal{D}^T_{(0)a} e^{\underline{i}}_{(0)b} & = & \nabla^T_{(0)a} e^{\underline{i}}_{(0)b}+\omega_{(0)a}^{\underline{i}}\tau_{(0)b}+\omega_{(0)a}{}^{\underline{i}}{}
_{\underline{j}}e^{\underline{j}}_{(0)b}\,,\label{eq:nablae}\\
\hspace{-.5cm}\mathcal{D}^T_{(0)a} v_{(0)}^b & = & \nabla^T_{(0)a}v_{(0)}^b+\omega_{(0)a}^{\underline{i}}e^b_{(0)\underline{i}}\,,\\
\hspace{-.5cm}\mathcal{D}^T_{(0)a} e^b_{(0)\underline{i}} & = & \nabla^T_{(0)a} e^b_{(0)\underline{i}}-\omega_{(0)a}{}^{\underline{j}}{}_{\underline{i}}e^b_{(0)\underline{j}}\,,\label{eq:covder4}
\end{eqnarray}
where $\nabla^T_{(0)a}$ contains some not necessarily symmetric connection $\Gamma^{Tc}_{(0)ab}$ that we assume to be invariant under local $SO(2)$ rotations and $\tau_{(0)c}\delta\Gamma_{(0)ab}^{Tc}=0$ under boosts so that $\delta(\mathcal{D}^T_{(0)a} \tau_{(0)b})=0$. The condition $\nabla^T_{(0)a}\tau_{(0)b}=0$ restricts $\Gamma^{Tc}_{(0)ab}$ but this condition will hold in all our cases. From demanding that $\delta(\mathcal{D}^T_{(0)a} e^{\underline{i}}_{(0)b})=\tau_{(0)b}\omega_{(0)}^c\mathcal{D}^T_{(0)a} e^{\underline{i}}_{(0)c}+\Lambda_{(0)\underline{j}}^{\underline{i}}\mathcal{D}^T_{(0)a} e^{\underline{j}}_{(0)b}$ we can derive the transformation properties of $\omega_{(0)a}{}^{\underline{i}}{}_{\underline{k}}$ and $\omega_{(0)a}^{\underline{i}}$ under boosts and $SO(2)$ transformations. We impose the vielbein postulate $\mathcal{D}^T_{(0)a}e^{\underline{i}}_{(0)b}=\mathcal{D}^T_{(0)a} v_{(0)}^b=\mathcal{D}^T_{(0)a} e^b_{(0)\underline{i}}=0$ to relate the boost $\omega_{(0)a}^{\underline{i}}$ and $SO(2)$ connection $\omega_{(0)a}{}^{\underline{i}}{}_{\underline{k}}$ to the affine connection $\Gamma^{Tc}_{(0)ab}$.

We will choose the symmetric part of $\Gamma_{(0)ab}^{Tc}$ to take the same functional form as it does for Newton--Cartan geometry \cite{Dautcourt,Andringa:2010it}
\begin{eqnarray}
\Gamma_{(0)ab}^c & = & -\frac{1}{2}v_{(0)}^c\left(\partial_a\tau_{(0)b}+\partial_b\tau_{(0)a}\right)\nonumber\\
&&+\frac{1}{2}\Pi_{(0)}^{cd}\left(\partial_a \Pi_{(0)bd}+\partial_b \Pi_{(0)ad}-\partial_d \Pi_{(0)ab}\right)\nonumber\\
&&-\frac{1}{2}\Pi_{(0)}^{cd}\left(F_{(0)da}\tau_{(0)b}+F_{(0)db}\tau_{(0)a}\right)\,,\label{eq:Gamma0}
\end{eqnarray}
with $F_{(0)ab}=\partial_a A_{(0)b}-\partial_b A_{(0)a}$ and $\Pi_{(0)ab}=\delta_{\underline{i}\underline{j}}e^{\underline{i}}_{(0)a}e^{\underline{j}}_{(0)b}$. This choice is naturally suggested by the null reduction \cite{Duval:1984cj,Julia:1994bs} of the 5D boundary metric.

Equation \eqref{eq:nablatau} then implies that the torsion is
\begin{equation}
T_{(0)ab}^c=-\frac{1}{2}v_{(0)}^c\left(\partial_a\tau_{(0)b}-\partial_b\tau_{(0)a}\right)\,. \label{torsiontensor}
\end{equation}
These two choices define what we mean by torsional Newton--Cartan geometry (TNC). It has the property that the degenerate metrics satisfy $\nabla^T_{(0)a}\tau_{(0)b} = 0$ and $\nabla^T_{(0)a}\Pi_{(0)}^{bc} = 0$. An important special case is when $\tau_{(0)a}$ is hypersurface orthogonal, i.e. $\tau_{(0)[a}\partial_{b}\tau_{(0)c]}=0$, that we refer to as temporal or twistless Newton--Cartan geometry (TTNC).

\noindent\textbf{Boundary Conditions.} We distinguish three types of asymptotically Lifshitz space-times depending on the behavior of $\tau_{(0)a}$. When $d\tau_{(0)}=0$ the boundary geometry is ordinary Newton--Cartan, i.e. no torsion. In this case there is an asymptotic scale symmetry and locally $\tau_{(0)a}=\partial_a t$ where $t$ is absolute time. We call these solutions asymptotically Lifshitz (ALif). When $\tau_{(0)}\wedge d\tau_{(0)}=0$ we call, following \cite{Ross:2011gu}, the solutions asymptotically locally Lifshitz (AlLif). In this case $\tau_{(0)a}$ defines hypersurfaces of absolute simultaneity but there is no absolute time. Finally, since the equations of motion also admit solutions with $\tau_{(0)}\wedge d\tau_{(0)}\neq 0$ we refer to these as the Lifshitz UV (see table \ref{table1}).
\begin{table}[h!]
      \centering
      \begin{tabular}{|c|c|c|c|}
      \hline
Asymptotics & $\tau_{(0)}\wedge d\tau_{(0)}$ & $d\tau_{(0)}$ & Boundary Geometry\\
             \hline
             ALif & $0$ & $0$ & NC\\
             \hline
             AlLif & $0$ & $\neq 0$ & TTNC\\
            \hline
            Lif UV & $\neq 0$ & $\neq 0$ & TNC\\
             \hline
      \end{tabular}
      \caption{Indicated are the 3 different boundary conditions. The last column indicates the type of boundary geometry.}\label{table1}
\end{table}

\noindent\textbf{Vevs and Ward identities.} 

Computing the variation of the renormalized 4D on-shell action, obtained from \eqref{eq:actionmodel} by 
our dimensional reduction, we find
\begin{align}
& \delta S_{\text{ren}}^{\text{os}} = - \frac{2\pi L}{\kappa_{5}^{2}}\int_{\partial \mathcal{M}}d^{3}x e_{(0)}\left(-S^{\underline{t}}_{(0)a}\delta v^{a}_{(0)}+S^{\underline{i}}_{(0)a}\delta e^{a}_{(0)\underline{i}}
\right.\nonumber\\
& \left.+T_{(0)}^{\underline{t}}\delta A_{(0)\underline{t}}+ T^{\underline{i}}_{(0)}\delta A_{(0)\underline{i}} +\langle\mathcal{O}_\chi\rangle\delta\chi_{(0)}+ \langle\mathcal{O}_\phi\rangle \delta \phi_{(0)}\right.\nonumber\\
 &\left.+\frac{1}{2}\left(S_{(0)\underline{t}}^{\underline{t}}-S_{(0)\underline{i}}^{\underline{i}}+A_{(0)\underline{t}}T_{(0)}^{\underline{t}}-A_{(0)\underline{i}}T_{(0)}^{\underline{i}}+2\langle\mathcal{O}_\Phi\rangle\right)\delta\Phi_{(0)} \right.\nonumber\\
 &\left. - \mathcal{A}_{(0)}\frac{\delta r}{r}\right)\,,\label{eq:onshellvar}
 \end{align}
where we varied with respect to the inverse vielbeins. The term proportional to $\delta\Phi_{(0)}$ contains contributions involving other vevs due to the appearance of $\Phi_{(0)}$ in \eqref{etexp}--\eqref{eiexp}. By reducing the variation of the 5D action \eqref{eq:actionmodel} and going on-shell we obtain the following relation between 5D and 4D vevs
\begin{eqnarray}
\hspace{-.3cm}e_{(0)}^{\underline{i}\,b}\hat t_{ab} & = & S^{\underline{i}}_{(0)a}+A_{(0)}^{\underline{i}}S^{\underline{t}}_{(0)a}\,,\quad \hat t_{au} = S^{\underline{t}}_{(0)a}\,,\label{eq:hat t ab in 4D vevs}\\
\hspace{-.3cm}\hat t_{uu} & = & -T_{(0)}^{\underline{t}} \label{eq:tuu}\,,\quad \langle\hat O_{\hat\phi}\rangle  = \langle O_\phi\rangle\,,\quad\langle\hat O_{\hat\chi}\rangle=\langle O_\chi\rangle \,,\label{eq:4D in 5D vevs 7}
\end{eqnarray}
together with the relations
\begin{eqnarray}
\hspace{-.3cm}0 & = & S_{(0)\underline{t}}^{\underline{t}}-S_{(0)\underline{i}}^{\underline{i}}+A_{(0)\underline{t}}T_{(0)}^{\underline{t}}-A_{(0)\underline{i}}T_{(0)}^{\underline{i}}+2\langle O_\Phi\rangle\,,\label{eq:vevrelation}\\
\hspace{-.3cm}0 & = & A_{(0)}^{\underline{i}}T_{(0)}^{\underline{t}}+e_{(0)}^{\underline{i}\,a}S_{(0)a}^{\underline{t}}+T_{(0)}^{\underline{i}} \,,\label{eq:vevrelation1}\\
\hspace{-.3cm}0 & = & S_{(0)}^{\underline{i}\underline{j}}+A_{(0)}^{\underline{i}}S_{(0)a}^{\underline{t}\underline{j}}-(\underline{i}\leftrightarrow\underline{j})\,,\label{eq:vevrelation2}
\end{eqnarray}
where $S_{(0)}^{\underline{i}\underline{j}}=e_{(0)}^{\underline{j}\,a}S_{(0)a}^{\underline{i}}$ and $S_{(0)}^{\underline{t}\underline{j}}=e_{(0)}^{\underline{j}\,a}S_{(0)a}^{\underline{t}}$. 

Equations \eqref{eq:vevrelation1} and \eqref{eq:vevrelation2} are the Ward identities associated with local boosts (under boosts the boundary gauge field transforms as $\delta A_{(0)a}=-\Lambda_{(0)a}$ as follows by relating the 4- to the 5-dimensional sources) and $SO(2)$ rotations. Relation \eqref{eq:vevrelation} is a Ward identity associated with the local dilatation $\delta\Phi_{(0)} = \Lambda_{(0)}$. By going to 5 dimensions this transformation can be viewed as a local dilatation leaving the asymptotically locally AdS boundary metric with $\hat h_{(0)uu}=0$ invariant. The transformation $\delta\Phi_{(0)} = \Lambda_{(0)}$ is only a leading order symmetry of the full FG expansion, which is enough to derive a Ward identity. By substituting \eqref{eq:vevrelation} in \eqref{eq:onshellvar} the variation of the remaining sources become unconstrained.

We derive the remaining Ward identities by reducing the 5D Penrose--Brown--Henneaux transformations \cite{Penrose:1986ca,Brown:1986nw} that leave the form of the FG expansion form invariant. This leads to three sets of local symmetries: boundary diffeomorphisms obtained by transforming the sources as boundary tensors, gauge transformations
\begin{equation}
\delta A_{(0)a}=\partial_a\Sigma_{(0)}\,,\qquad\delta\chi_{(0)}=k\Sigma_{(0)}
\end{equation}
and anisotropic Weyl transformations \cite{Horava:2009vy}
\begin{equation}
\delta X_{(0)}=\lambda \xi^r_{(0)}X_{(0)}\,,
\end{equation}
where $X_{(0)}$ is any of the sources listed in the first row of table \ref{table:scalingdimensions} and $\lambda$ is the weight.
\begin{table}[h!]
      \centering
      \begin{tabular}{|c|c|c|c|c|c|c|c|}
      \hline
 &  $v^{a}_{(0)}$ & $e^{a}_{(0)\underline{i}}$ & $A_{(0)\underline{t}}$ & $A_{(0)\underline{i}}$ &$\Phi_{(0)}$ & $\phi_{(0)}$ & $\chi_{(0)}$\\
  \hline
 $\lambda$ &2&1&2&1&0&0&0    \\
         \hline
          & $S_{(0)a}^{\underline{t}}$ & $S_{(0)a}^{\underline{i}}$ & $T_{(0)}^{\underline{t}}$ & $T_{(0)}^{\underline{i}}$ &$\langle O_\Phi\rangle$ & $\langle O_\phi\rangle$ & $\langle O_\chi\rangle$\\
  \hline
 $4-\lambda$ &2&3&2&3&4&4&4    \\
 \hline
           \end{tabular}
      \caption{Scaling dimensions of the 4D sources and vevs.}\label{table:scalingdimensions}
\end{table}

To obtain the transformation properties of the 4D vevs we use \eqref{eq:hat t ab in 4D vevs}--\eqref{eq:4D in 5D vevs 7} together with the PBH transformation of $\hat t_{\hat a\hat b}$. The boundary stress-energy tensor is 
\begin{equation}
\mathcal{T}_{(0)a}^b=S_{(0)a}^b+T_{(0)}^b\frac{1}{k}\partial_a\chi_{(0)}\,, \label{Tshifted}
\end{equation}
which is gauge invariant (see also \cite{Hollands:2005ya,Ross:2009ar}). The scaling dimensions (obtained by taking $\xi^r_{(0)}$ constant) for the 4D vevs are listed in the second row of table \ref{table:scalingdimensions}. 

The boundary covariant form of the Ward identities for gauge, anisotropic Weyl and diffeomorphism invariance are then given by
\begin{eqnarray}
k\left\langle O_\chi\right\rangle & = & e_{(0)}^{-1}\partial_a\left(e_{(0)}T_{(0)}^a\right)\,,\\
\mathcal{A}_{(0)} & = & 2\mathcal{T}_{(0)\underline{t}}^{\underline{t}}+2B_{(0)\underline{t}}T_{(0)}^{\underline{t}}+\mathcal{T}_{(0)\underline{i}}^{\underline{i}}+B_{(0)\underline{i}}T_{(0)}^{\underline{i}}\,,\label{eq:traceWardidentity}\\
\nabla^T_{(0)b}\mathcal{T}_{(0)a}^b & = & -\mathcal{T}_{(0)b}^c\left(-\tau_{(0)c}\nabla^T_{(0)a}v_{(0)}^b+e_{(0)c}^{\underline{i}}\nabla^T_{(0)a}e_{(0)\underline{i}}^b\right)\nonumber\\
&&+2T_{(0)ac}^b\mathcal{T}_{(0)b}^c+2T_{(0)bc}^b\mathcal{T}_{(0)a}^c-T_{(0)}^{\underline{t}}\partial_aB_{(0)\underline{t}}\nonumber\\
&&-T_{(0)}^{\underline{i}}\partial_aB_{(0)\underline{i}}-\langle O_\phi\rangle\partial_a\phi_{(0)}\,,\label{eq:LifshitzdiffeoWardtorsion}
\end{eqnarray}
where $\nabla_{(0)a}^T$ is defined just below \eqref{eq:covder4} and where $B_{(0)a} = A_{(0)a}-\frac{1}{k}\partial_a\chi_{(0)}$ is the boundary massive vector field. Using a TNC analogue notion of boundary conformal Killing vectors it is possible to define boundary conserved currents \cite{Christensen:2013rfa}. The anisotropic Weyl anomaly $\mathcal{A}_{(0)}$ takes the form of a Horava--Lifshitz action on TNC (as opposed to a Lorentzian geometry) \cite{Christensen:2013rfa}. Counting all sources and vevs and subtracting local symmetries/Ward identities we end up with 6+6 sources and vevs.

\noindent\textbf{A Second UV Completion.} The 4D FG expansion contains the extra free function $v_{(0)}^av_{(0)}^b\hat t_{ab}$ which does not appear in \eqref{eq:hat t ab in 4D vevs}. This is the vev of an irrelevant operator $\hat h_{(0)uu}$ that had to be switched off in order to have a $z=2$ Lifshitz UV. Switching it on modifies the UV but not the IR as can be seen from the solution
\begin{eqnarray}
ds^2 & = & \frac{1}{r^2}e^\Phi\left[-\left(1+\frac{k^2}{4}g_s^2r^2\right)^{-1}dt^2+dx^2+dy^2+dr^2\right]\,,\nonumber\\
e^{2\Phi} & = & r^{-2}\left(1+\frac{k^2}{4}g_s^2r^2\right)\,,\; A = \left(1+\frac{k^2}{4}g_s^2r^2\right)^{-1}\!dt\,,\label{eq:hyperscalingUV3}
\end{eqnarray}
with $\chi$ and $\phi=\log g_s$ constant. For small $r$ (UV) this is a $\theta=-1$ and $z=1$ hyperscaling geometry and for large $r$ (IR) this is our $z=2$ Lifshitz geometry. Expanding around $r=\infty$ an irrelevant mode with fall off $r^{-2}$ appears. The 5D uplift has $\hat h_{(0)uu}>0$.

\noindent\textbf{Discussion.} We expect the TNC boundary geometry to be universal in Lifshitz holography for all values of $z>1$ since the argument of the contraction of the bulk local Lorentz group works for any $z>1$. Further, many other features of our setup, such as the role of the boundary gauge field and the benefit of using a non-radial gauge could play an important role in the development of Lifshitz holography. Finally, we expect these results to be relevant for a fluid/gravity type derivation of Lifshitz hydrodynamics \cite{Hoyos:2013qna} which has potential applications to holographic realizations of Son's model for the effective theory of the fractional quantum Hall effect that relies on Newton--Cartan geometry \cite{Son:2013rqa}.

\noindent\textbf{Acknowledgements.}
We thank Marco Baggio, Matthias Blau, Jan de Boer, Wissam Chemissany, Kristian Holsheimer, Elias Kiritsis, Ioannis Papadimitriou, Simon Ross, Marika Taylor and L\'arus Thorlacius for useful discussions. The work of JH and NO is supported in part by the Danish National Research Foundation project ``Black holes and their role in quantum gravity''. BR acknowledges support from the Swiss National Science Foundation through the fellowship PBBEP2\_144805.

\addcontentsline{toc}{section}{References}

\bibliography{LifshitzHR-letter}

\end{document}